\documentstyle[epsfig,aps,prl,twocolumn]{revtex}
\begin{document}

\wideabs{

\title{Raman spectrum and lattice parameters of MgB$_2$
as a function of pressure}

\author{Alexander F. Goncharov, Viktor V. Struzhkin, Eugene Gregoryanz,
Jingzhu Hu, Russell J. Hemley, Ho-kwang Mao,}

\address{Geophysical Laboratory and Center for High Pressure
Research, Carnegie Institution of Washington, \\
5251 Broad Branch Road NW, Washington D.C. 20015 U.S.A}

\author{G. Lapertot \cite{a1}, S. L. Bud'ko, P. C. Canfield}

\address{Ames Laboratory and Department of Physics and Astronomy,
Iowa State University, Ames, IA 50010}


\maketitle

\begin{abstract}

We report Raman spectra and synchrotron x-ray diffraction
measurements of lattice parameters of polycrystalline MgB$_2$
under hydrostatic pressure conditions up to 15 GPa.
An anomalously broadened Raman band at 620 cm$^{-1}$ is observed that exhibits a large linear pressure shift of its frequency.
The large mode damping and Gr\"uneisen parameter indicate a highly
anharmonic nature of the mode, broadly consistent with theoretical
predictions for the E$_{2g}$ in-plane boron stretching mode.
The results obtained may provide additional constraints on the
electron-phonon coupling in the system.

\end{abstract}
}

The newly discovered \cite{Akimitsu} high-temperature superconductor
MgB$_2$ has attracted considerable interest from theoretical and experimental
points of view. Theory indicates that MgB$_2$ can be treated as phonon
mediated superconductor with very strong coupling
\cite{kortus,an,kong,yildirim}. Calculations show that the strongest coupling is realized for the near-zone center in-plane optical phonon (E$_{2g}$ symmetry) related to vibrations of the B atoms \cite{an,kong,yildirim}.
According to recent calculations, this phonon is very anharmonic because
of its strong coupling to the partially occupied planar B $\sigma$ bands
near the Fermi surface \cite{yildirim}. The frequency of this
phonon ranges from 460 to 660 cm$^{-1}$ according to different
computation techniques \cite{kortus,an,kong,yildirim,satta}.
The phonon density of states for MgB$_2$ has been determined by
neutron inelastic scattering \cite{yildirim,osborn,sato},
but the E$_{2g}$ mode could not be detected separately.
Raman experiment \cite{bohnen} indicated a presence of a broad
mode at 72 meV (580 cm$^{-1}$) in agreement with calculations
for the E$_{2g}$ mode. Transport, magnetic susceptibility, and
specific heat measurements show a large isotope effect consistent with
phonon mediated superconductivity \cite{bud'ko}.

Pressure is an important variable, that can be used to tune
physical properties and compare the results with theoretical predictions.
Pressure effects on superconductivity studied to 1.84 GPa
\cite{lorenz} and 0.5 GPa \cite{tomita} show a
decrease of $T_c$  with the rate of 1.6 K/GPa and 1.11 K/GPa,
respectively (see also Ref. \cite{monteverde}).
Compressibility data have been obtained by neutron diffraction (to 0.62 GPa)
\cite{jorgensen} and synchrotron x-ray diffraction (to 6.15 GPa
\cite{prassides} and 8 GPa \cite{vogt}).  Based on theoretical calculations
of the electronic density of states at the Fermi level, which show a very
moderate decrease with pressure, the dominant contribution to the decrease of
$T_c$ under pressure has been proposed to be due to an increase in phonon
frequency \cite{loa}.

In this Letter we present Raman measurements of the phonon
mode including the first measured under pressure. We find that the
E$_{2g}$ band is unusually broaden and shows a large positive pressure shift
of the frequency.  We also present x-ray diffraction data, which allow us to
determine lattice parameters on the sample from the same batch in purely
hydrostatic conditions to 12 GPa. As a result, we determined the E$_{2g}$
mode Gr\"uneisen parameter, which is much larger than in case of "normal"
materials. We ascribe this to large anharmonic effects predicted by theory.
The increase in phonon frequency measured in this work explains the reported
T$_C$ drop with pressure.

Samples of Mg$^{10}$B$_2$ were similar to those used in Refs.
\cite{bud'ko,finnemore}.  They are essentially in a powdered form
consisting of aggregates of 30-50 $\mu$m linear dimensions, which is ideal
for high-pressure experiments.  Our experiments have been done with various
types of diamond anvil cells.  In Raman experiments a long piston-cylinder
cell was used and Ne served as a pressure transmitting medium \cite{neon}.
Synthetic ultrapure diamonds were used as anvils to reduce background
fluorescence.  Raman scattering was excited in a 145$^\circ$ geometry (see
Ref.  \cite{iron}) to reduce further background from diamond Raman and
that originating from spuriously reflected elastic light. The
spectra were recorded with a single-stage spectrograph equipped
with a CCD detector and holographic notch filters (150-5000 cm$^{-1}$),
although occasional measurements were also done with a conventional
triple spectrometer to cover the lower frequency range. X-ray diffraction
was measured with a two-side open diamond cell in an
energy-dispersive configuration at X17C beamline of the
National Synchrotron Light Source with 2$\theta$=10$^\circ$ \cite{airapt}.
In the x-ray experiment we used helium as a pressure transmitting medium,
which is purely hydrostatic to 12 GPa.  Pressure was determined by the
standard ruby fluorescence technique.  All measurements were performed at
room temperature.

\begin{figure}
\centerline{\epsfig{file=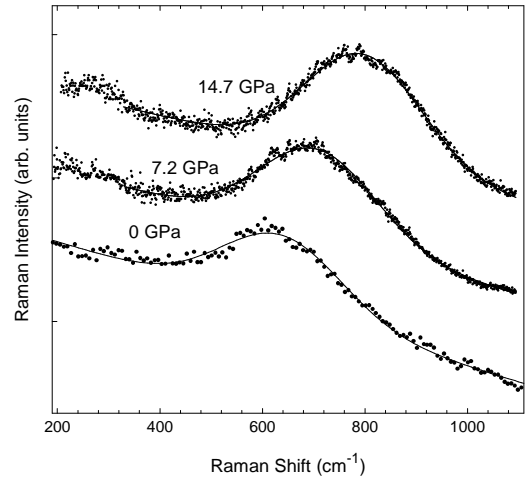,width=10cm}}
\caption{Raman spectra of MgB$_2$ at elevated pressures.
Spectra are shifted vertically for clarity. Points are
experimental data and solid lines represent the phenomenological
fits (see text) to the spectra in the appropriate spectral range.
The excitation wavelength is 514.5 nm.}
\label{fig1}
\end{figure}

\begin{figure}
\centerline{\epsfig{file=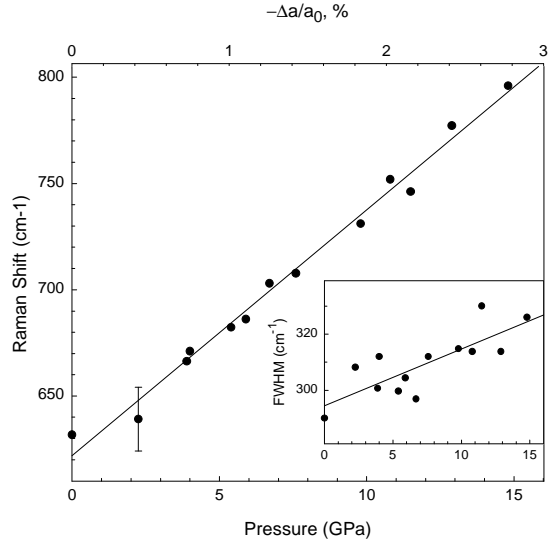,width=10cm}}
\caption{Raman frequency as a function of pressure and
relative compression of the latice parameter $a$ axis (upper scale). Points
are experimental data of the frequencies determined from the
phenomenological fits of the spectra.
The solid line is a linear fit. Inset shows the pressure dependence
of the damping obtained by the same fitting procedure.
}
\label{fig2}
\end{figure}

Figure \ref{fig1} presents the Raman spectra at different pressures.
The broad band observed is a Raman excitation as shown
by changing the excitation wavelength and by anti-Stokes measurements.
It has also been checked that the signal originates from MgB$_2$
because identical spectra were recorded by separate micro Raman
measurements from individual micron-size grains (shown in Fig. \ref{fig1}
as the 0 GPa spectrum). Also, the Raman spectra contain a wide unstructured
background component (presumably of electronic origin as in the cuprate HTSC
materials \cite{blumberg}), which increases intensity at lower frequencies.
Pressure leads to an increase in the frequency of the broad band
without any appreciable change
of its shape. The spectra can be fitted reasonably well with
a combination of a linear background and a Gaussian peak. The
frequency determined by this procedure is plotted as a function
of pressure in Fig. \ref{fig2}. The pressure dependence is linear within the
accuracy of the experiment. No essential pressure dependence
of the mode damping was found (inset in Fig. \ref{fig2}).

Factor-group analysis predicts for MgB$_2$ (space group $P6/mmm$, Z=1)
B$_{1g}$ +E$_{2g}$+ A$_{2u}$+E$_{2u}$ zone center optical modes,
of which only E$_{2g}$ is Raman active. Thus, it is natural to assign
the band observed at 620 cm$^{-1}$ at ambient conditions to the E$_{2g}$ mode
(see also Ref. \cite{bohnen}).
The experimental frequency agrees well with theoretical calculations
\cite{yildirim,bohnen}. The anomalously large
linewidth (FWHM=300 cm$^{-1}$) can be ascribed to
large electron-phonon coupling \cite{bohnen}, which will be described below.

\begin{figure}
\centerline{\epsfig{file=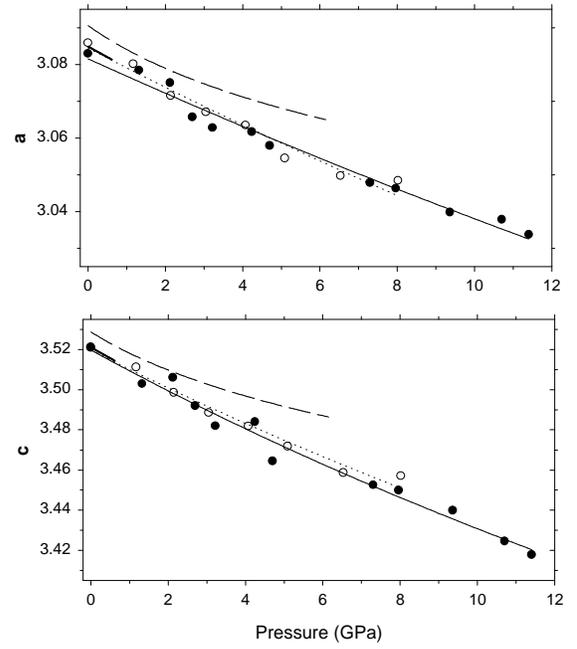,width=10cm}}
\caption{Experimental pressure dependences of the lattice parameters.
Filled circles with solid line (Murnagan fit) are our data;
thick solid lines are from Ref. \protect\cite{jorgensen};
dashed lines are from Ref. \protect\cite{prassides};
open circles and dotted lines are from Ref. \protect\cite{vogt}.
}
\label{fig3}
\end{figure}

The experimental pressure dependencies of lattice parameters determined
by x-ray diffraction are shown in Fig. \ref{fig3}. Our data are in
good agreement with Refs. \cite{jorgensen,vogt}, while the results of
Ref. \cite{prassides} show
systematically larger lattice parameters and yet comparable compressibility.
We calculated the bulk modulus K$_{0}$ assuming "normal" behavior and
K$_{0}^{\prime}$=4, which is typical for covalent and metallic bonding
\cite{review}(our data do not allow us to fit data with two parameters
K$_{0}$ and K$_{0}^{\prime}$). The result is 155(10) GPa in good agreement
with Refs.\cite{jorgensen,vogt,loa}. Similar calculations for in-plane and
out of plane compressibilities give $\beta_{a}$=0.0016(2) GPa$^{-1}$ and
$\beta_{c}$=0.0030(2) GPa$^{-1}$.

Thus, the mode G\"runeisen parameter $\gamma$=K$_0$dln$\nu$/dP determined
from our data equals 2.9$\pm$0.3. In the case of anisotropic crystals
it would be more appropriate to scale the frequency shift of in-plane
mode with the variation of interatomic bond distance or lattice
parameter $a$ \cite{hanfland}. The corresponding component of the
Gr\"uneisen parameter ($\gamma$=dln$\nu$/3dln$a$) is 3.9$\pm$0.4.
These values are substantially larger than those expected for
the phonon in a compound with covalent bonding \cite{sherman},
which should be dominant for this mode, where typically, $\gamma$ does
not exceed 2. For example, for graphite $\gamma$=1.06 \cite{hanfland}
and for iron (with metallic bonding partially present in our
case) $\gamma$=1.7 \cite{iron}. Larger $\gamma$'s are normally
related to increased anharmonicity of the particular normal vibration
\cite{zallen}. It can also be a consequence of a soft mode behavior
when the system is approaching (or departing from) a structural instability
(e.g. Ref. \cite{olijnyk}).

The proposed assignment of the Raman peak observed to the first-order 
phonon scattering is not the only possible.  Alternatively, if the first 
order scattering is inherently weak, the observed Raman peak can in principle 
be second order (e.g., due to overtones and combinations of the zone boundary 
acoustical phonons). However, this interpretation does not seem plausible 
because no higher frequency peak corresponding to combinations of acoustic 
and optical or two optical modes is observed.  Also, the observed excitation 
may not necessarily be of phonon nature, but in principle could be a magnetic 
excitation \cite{furukawa} (e.g. two-magnon peak, which is strongly dependent 
on interatomic distances with $\gamma$=3.5 \cite{massey}). However, we 
believe that the data available strongly suggest the first-order phonon 
interpretation because of the agreement with the calculated
frequency \cite{yildirim,bohnen} and the linewidth  \cite{bohnen}.

Theoretical calculations \cite{yildirim} suggest a scenario
with the E$_{2g}$ phonon strongly coupled to electronic excitations.
Our data show a very broad, strongly pressure dependent excitation, which is
consistent with this idea.  Conventional anharmonicity (not coupled to
electronic degrees of freedom) is expected to exhibit some variation with
pressure (see e.g.  Ref.  \cite{jayaraman}), which is not the case. Within
this picture, our data favor the coupling of the E$_{2g}$ phonon to the
electronic subsystem.

Finally, we address the observed strong pressure dependence of $T_c$
\cite{lorenz,tomita,monteverde}. Assuming a pressure independent density of
electronic states N(0), the averaged electron-ion matrix element $I$ and the
Coulomb pseudopotential $\mu^{\ast}$, one can get d$T_c$/dP=-1.5 K/GPa
with physically reasonable values of $\mu^{\ast}$=0.005-0.1 and
electron-phonon coupling constant $\lambda$=0.65-1
\cite{kortus,an,kong,yildirim,loa}. Thus, the pressure
dependence of $T_c$ can be easily explained by an increase in phonon
frequency as proposed in Refs. \cite{loa}.

In conclusion, we observed a strongly broadened Raman band of MgB$_2$
that shows anomalously large pressure dependence of its frequency.
This band and its pressure dependence can be interpreted as the
E$_{2g}$ zone center phonon, which is strongly anharmonic because
of coupling to electronic excitations.

We acknowledge financial support of CIW, NSLS, NSF and Keck Foundation.
Work at Ames laboratory was supported by the Director for Energy Research,
Office of Basic Energy Sciences, U.S. Department of Energy.



\end{document}